\documentclass[a4paper,11pt]{article}
\pdfoutput=1 

\usepackage{jinstpub}
\usepackage{subfigure}

\newcommand{\meg}{\ifmmode\mu \to {\rm e} \gamma\else$\mu \to \mathrm{e} \gamma$\fi}
\newcommand{\megc}{\ifmmode\mu^+ \to {\rm e}^+ \gamma\else$\mu^+ \to \mathrm{e}^+ \gamma$\fi}
\newcommand{\michel}{\ifmmode\mu^+ \to {\rm e}^+ \nu\bar{\nu}\else$\mu^+ \to \mathrm{e}^+ \nu\bar{\nu}$\fi}
\newcommand{\radiative}{\ifmmode\mu^+ \to {\rm e}^+\nu\bar{\nu}\gamma \else$\mu^+ \to \mathrm{e}^+ \nu\bar{\nu}\gamma$\fi}

\newcommand*{\egamma}         {E_{\mathrm{\gamma}}}
\newcommand*{\positron}      {\mathrm{e^+}}
\newcommand*{\epositron}      {E_\mathrm{e^+}}
\newcommand*{\ppositron}      {p_\mathrm{e^+}}

\newcommand*{\tegamma}        {t_{\mathrm{e^+ \gamma}}}
\newcommand*{\Thetaegamma}    {\Theta_{\mathrm{e^+ \gamma}}}

\newcommand*{\thetae}         {\theta_\mathrm{e^+}}
\newcommand*{\phie}           {\phi_\mathrm{e^+}}

\newcommand*{\ypos}          {y_\mathrm{e^+}}
\newcommand*{\zpos}          {z_\mathrm{e^+}}
\newcommand*{\ugamma}         {u_{\gamma}}
\newcommand*{\vgamma}         {v_{\gamma}}
\newcommand*{\wgamma}         {w_{\gamma}}
\newcommand*{\BR}     { {\cal B} }

\title{The MEGII detector}
 
\author[a]{ P.W.~Cattaneo}

\affiliation[a]{INFN Pavia, Via Bassi 6, I-27100 Pavia, Italy}

\emailAdd{paolo.cattaneo@pv.infn.it}

\abstract{
We present a report of the MEG II experiment, the upgrade of MEG, whose goal is to search for the forbidden 
decay \megc\ with increased precision.
After having briefly reviewed the motivation for such a search and the current limit due to MEG, 
we present the conceptual design of the detector detailing for each subdetector
the motivations and the extent of the upgrade and the expected resolution improvements.
Novel subdetectors and calibration hardware are introduced.
We conclude with the expected sensitivity of the MEGII experiment.
}

\collaboration[c]{on behalf of the MEG II collaboration}

\proceeding{INSTR17: Instrumentation for Colliding Beam Physics \\
  27 February -- 3 March, 2017 \\
  Novosibirsk, Russia}

\begin{document}

\maketitle
%
\section{Introduction}

\begin{figure}
\centering
\includegraphics[width=1.0\textwidth]{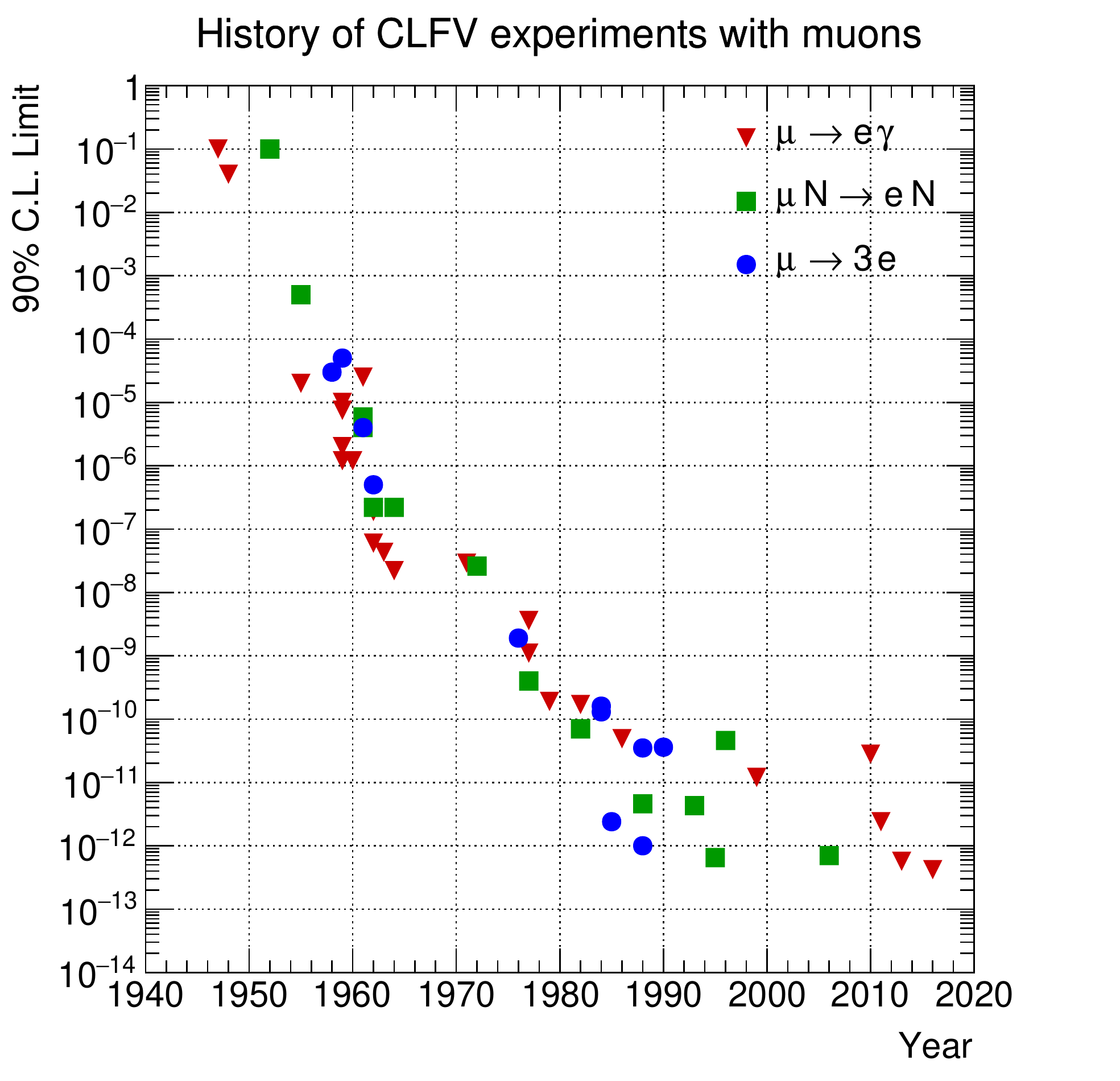}
\caption{Upper limits on cLFV processes as a function of the year. \label{fig:LFVlimits}}
\end{figure}

The experimental upper limits established in searching for cLFV processes with muons, including the \megc\ decay,
are shown in Fig.~\ref{fig:LFVlimits}.
\lq{}Surface\rq{} muon beams (i.e. beams of muons originating in the decay of $\pi^+$\rq{}s that stopped in the surface 
of the pion production target) with a monochromatic momentum of $\sim\!29~\mathrm{MeV}/c$,
offer the highest muon stop densities obtainable at present.

The MEG experiment \cite{megdet} at the Paul Scherrer Institute (PSI, Switzerland) used the world's most intense
beam with a stopping intensity of $3 \times 10^7~\mu^+/\mathrm{s}$ in the period 2009-2013. The signal of the possible 
two-body \megc\ decay at rest is distinguished from the background by measuring the photon energy $\egamma$,
the positron momentum $\ppositron$, their relative angle $\Thetaegamma$
and timing $\tegamma$ with the best possible resolutions.\footnote{In the following we will indicate the (1$\sigma$) resolution on a variable
with a $\Delta$ in front of that variable}

\section{Search for the \megc\ decay}

The \megc\ is practically forbidden in the Standard Model (SM). Even in presence of massive neutrinos, the SM predicts a
$\BR(\meg)$ below $10^{-50}$, which cannot be experimentally observed. Processes with charge Lepton Flavour Violation are 
therefore clean channels to look for possible new physics beyond the SM (BSM).
Many BSM models predict a measurable value of BR($\meg$) $\geq 10^{-14}$ ~\cite{barbieri1994,hisano-1999,Cali2006,Calibbi:2009wk,Calibbi:2012gr}.
On the basis of these theoretical predictions the MEG collaboration suggested \cite{meg1infn} to extend the sensibility of a \meg\ 
search to $\sim 10^{-13}$.

In the search of this decay positive muons at rest are stopped in a target. 
The kinematic of the signal events is a very simple double body decay with the energies of the $\positron$ and 
of the $\gamma$ equal to half of the rest muon mass $m_\mu$. The particles are emitted in opposite directions.
The background comes from radiative muon decay \radiative\ (RMD) close to the kinematic limit and, more relevant,
accidental coincidences from decays of different muons.

The MEG experiment took data in the years 2009-2013 \cite{megdet} improving the limit on this decay by almost a factor of 30,
down to BR($\meg$) $< 4.2\,10^{-13}$ \cite{meg2009,meg2010,meg2013,TheMEG:2016wtm}.
This result is presented in Fig.~\ref{fig:LFVlimits} together with the results of previous cLFV searches.

After the end of the run, the collaboration launched a redesign of the experiment, now called MEG~II, for further improving the
limit by almost an order of magnitude. The experiment has been redesigned, some parts refurbished, some designed from scratch, 
based on the experience acquired during the MEG run. After several years of R\&D and beam tests the collaboration is ready for 
data taking next year. We estimate that three full years of data taking are required to reach the design sensitivity of 
$\sim 5.\,10^{-14}$.

\section{The design of the MEG~II detector}

The MEG~II detector design is based on the MEG detector design \cite{megdet}. 
The MEG experiment exploits a surface $\mu^+$ high-intensity beam produced at the $\pi$E5 channel at PSI.
This beam is transported onto a thin (210~$\mu$m) stopping target located at the center of a superconducting magnet generating a gradient magnetic field.
This gradient field is shaped in such a way to sweep rapidly $\positron$ emitted at polar angle close to $90^\circ$ and to guarantee
bending radius weakly dependent on the polar angle emission.

The photon from the decay is detected by a large liquid xenon detector read out by PMTs measuring the energy, interaction time and position.
The positron momentum, direction and emission vertex on the target are measured by a set of drift chambers embedded in the magnetic field. 
The positron decay time is measured by two barrel shaped sets of scintillator bars read out by PMTs (TC)  \cite{DeGerone:2011te}.

\begin{figure*}
\centering
\includegraphics[width=1.0\textwidth]{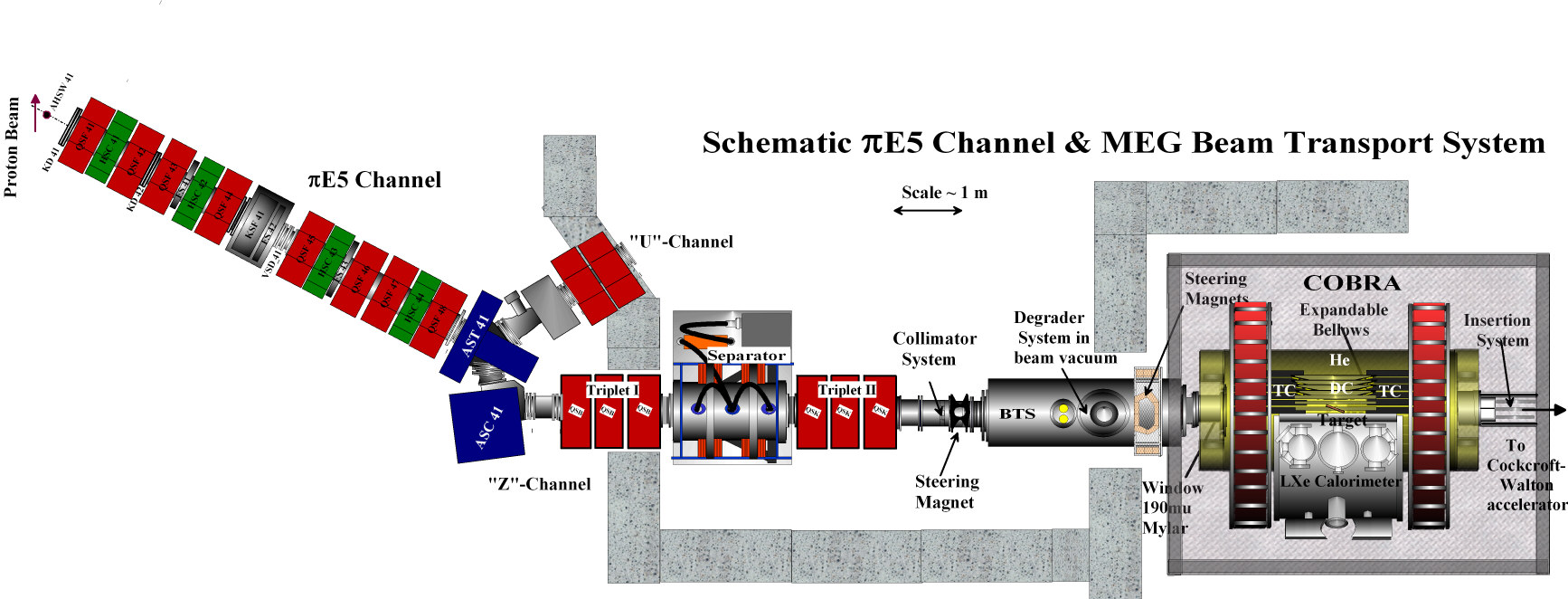}
\caption{MEG Beam line with the $\pi$E5 channel and MEG detector system incorporated in and around the COBRA magnet. \label{fig:beamline}}
\end{figure*}

The beam line, depicted in Fig.\ref{fig:beamline} as well as the COBRA magnet are retained in MEG~II, while a thinner target (140~$\mu$m)
has been selected to reduce multiple scattering and therefore the positron angular resolution.
The positron measuring part has been redesigned completely to overcome some of the problems that emerged during the MEG run.

\begin{figure}
\centering
\includegraphics[width=1.0\textwidth]{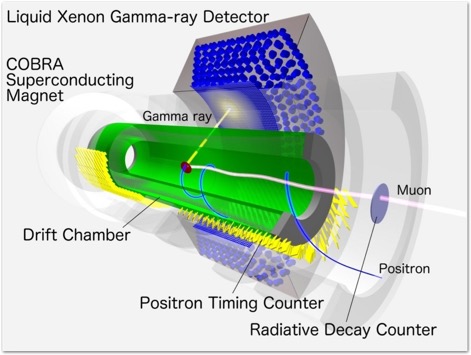}
\caption{A sketch of the MEG II experiment. The CDCH in green} 
\label{fig:megiisketch}
\end{figure}

\begin{figure}
\centering
\includegraphics[width=1.0\textwidth]{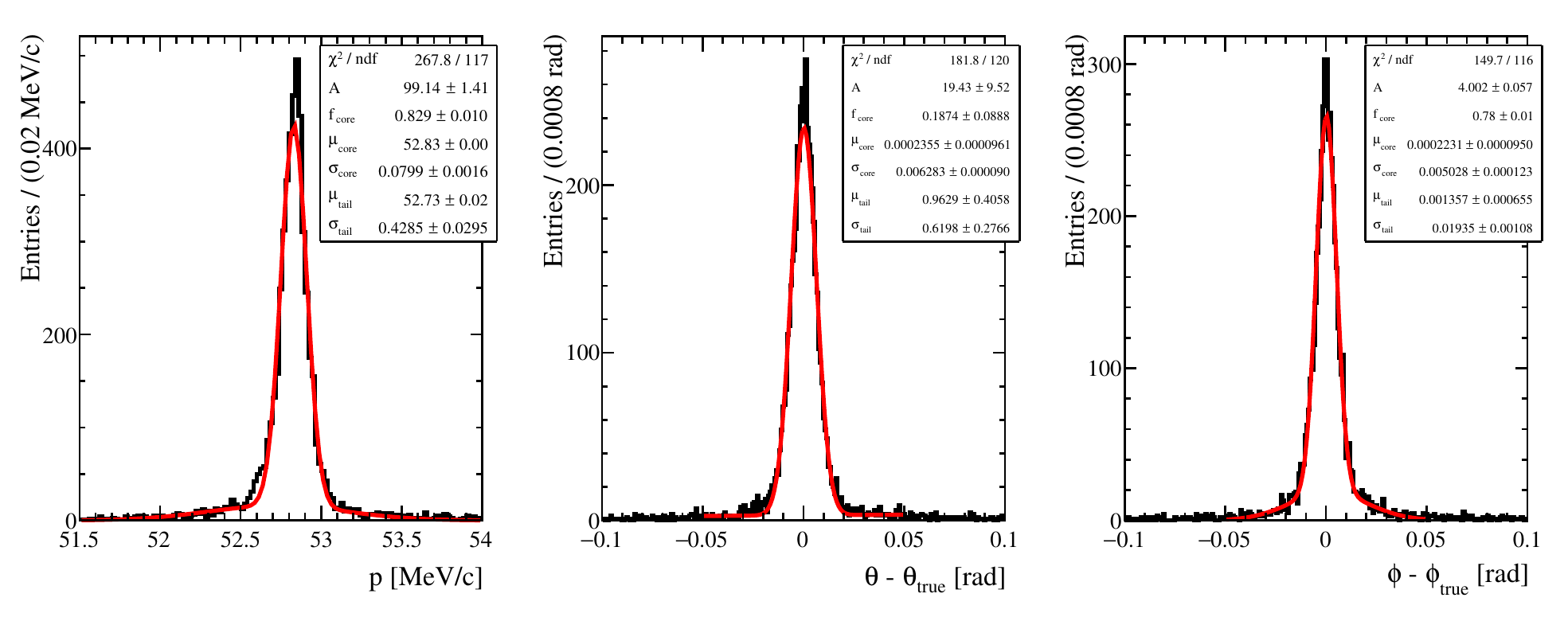}
\caption{Expected MEG~II CDCH momentum and angle resolutions, from a simulation having as inputs the results of the tests with prototypes.}
\label{fig:CDCHRes}
\end{figure}

The tracking detector in MEG`II is a single-volume Cylindrical Drift Chamber (CDCH), whose
design is optimized to satisfy the fundamental requirements of high
transparency and low multiple scattering contribution for 50~MeV positrons,
sustainable occupancy (at $\sim 7 \times 10^7$~ $\mu^+/$s stopped on target) and fast
electronics for cluster timing capabilities.
A sketch of the CDCH embedded in the MEGII detector is visible in Fig.\ref{fig:megiisketch} while the results of Monte Carlo 
simulations of the  momentum and angle resolutions are in Fig.~\ref{fig:CDCHRes}.

\begin{figure}
\centering
\includegraphics[width=1.0\textwidth]{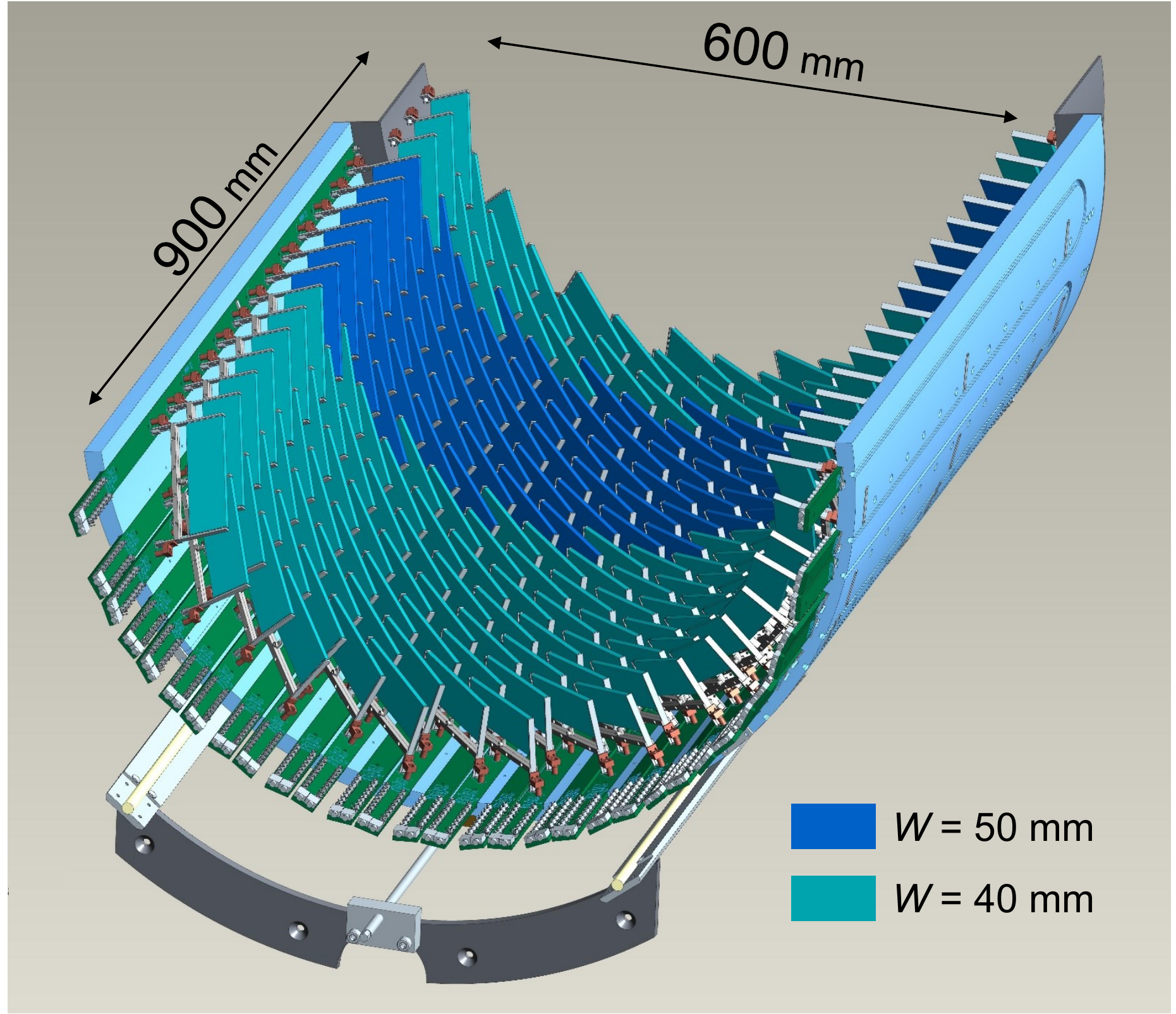}
\caption{Design of downstream pTC super-module.}
\label{fig:MEG-II-TC}
\end{figure}

\begin{figure*}[tb]
\centering
\includegraphics[width=1.0\linewidth]{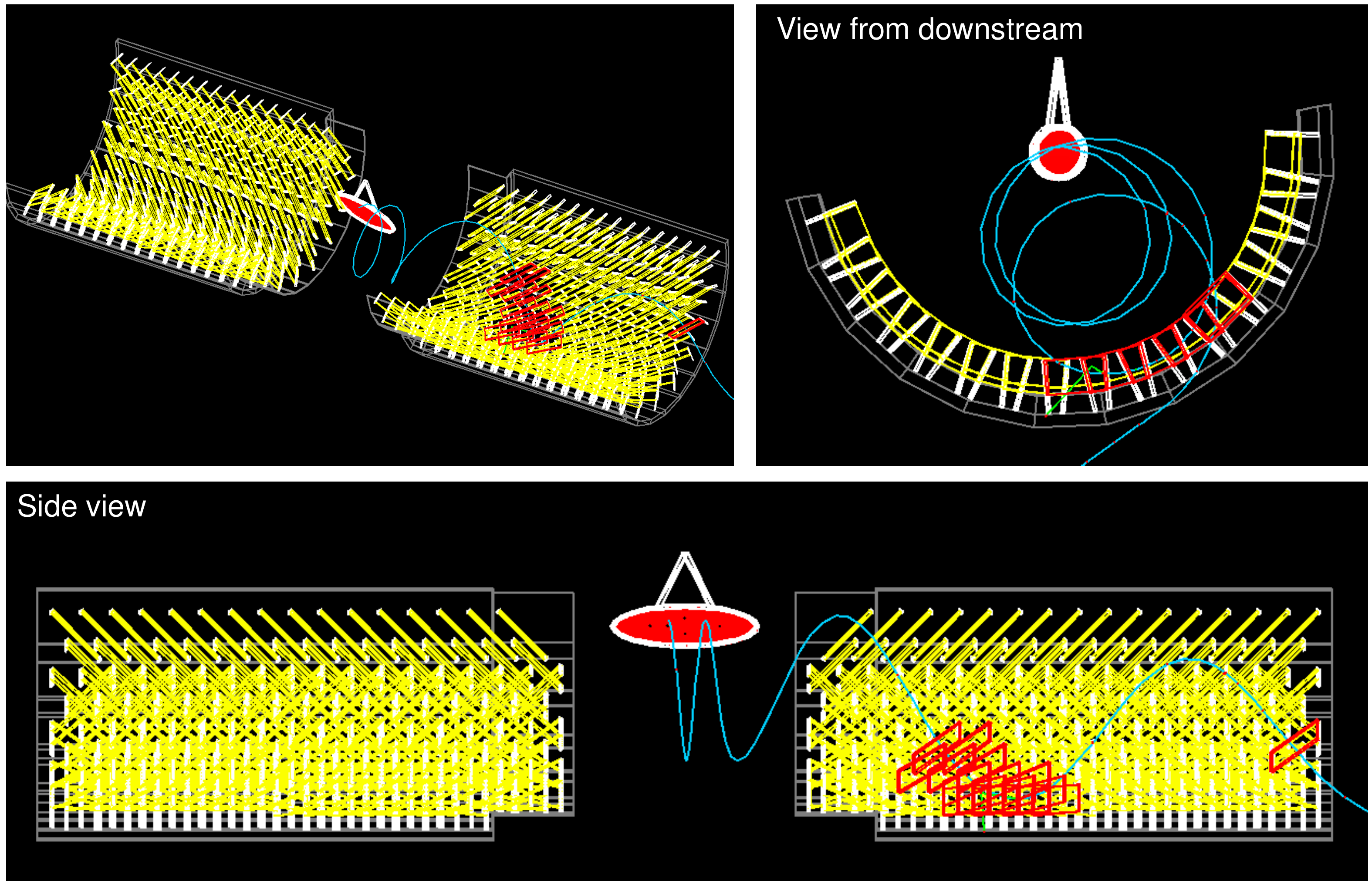}
\caption{An example of hit pattern by a simulated signal $\positron$. CDCH is not drawn in these figures.}
\label{fig:mcdisplay}
\end{figure*}

In MEG~II the TC is replaced by the pixelated Timing Counter (pTC). The pTC consists of two sets of scintillator pixels (256 each)
approximately barrel shaped, each pixel read out by two sets of Silicon PhotoMultipliers (SiPM) connected in parallel positioned
at opposite sides of the pixel \cite{ootani-nima,metcjinst2014,Cattaneo:2014uya,Nishimura:2015qev}. A design of the pTC is in 
Fig.~\ref{fig:MEG-II-TC} and an example of a simulated signal is in Fig.~\ref{fig:mcdisplay}.
An advantage of this configuration is that the positron timing is measured by many (in average $\sim 7$) pixels and the resolution
is improved by averaging down to $\sim 30$ ps. This resolution has been obtained in beam tests. 

\begin{figure}[b]
\centering
\includegraphics[width=1.0\textwidth]{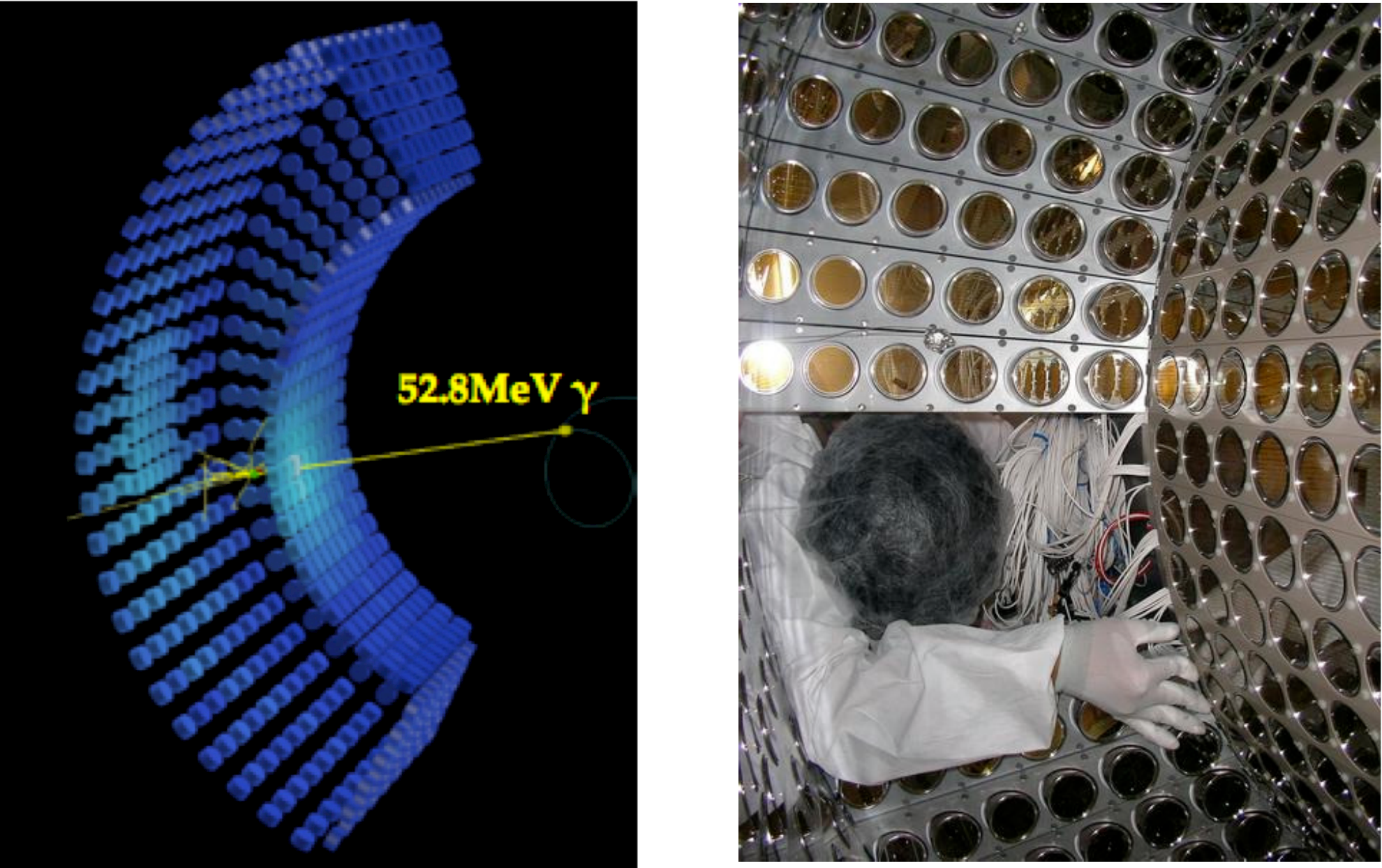}
\caption{\label{fig:LXeDetector}%
   MEG LXe photon detector with 900~$\ell$ LXe surrounded by 846 UV-sensitive PMTs.
}
\end{figure}

\begin{figure*}[tb]
\centering
\includegraphics[width=0.45\textwidth]{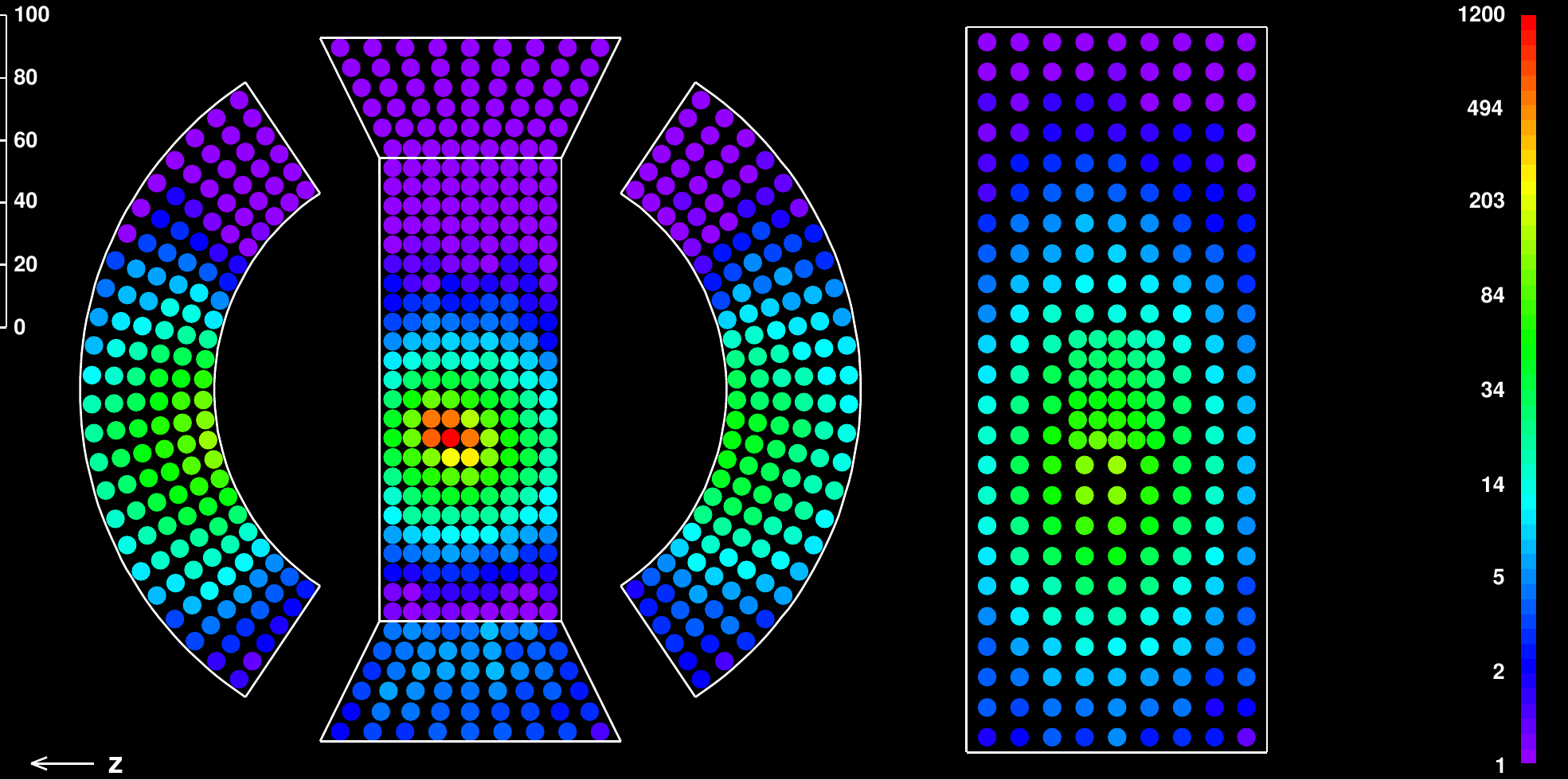}
\includegraphics[width=0.45\textwidth]{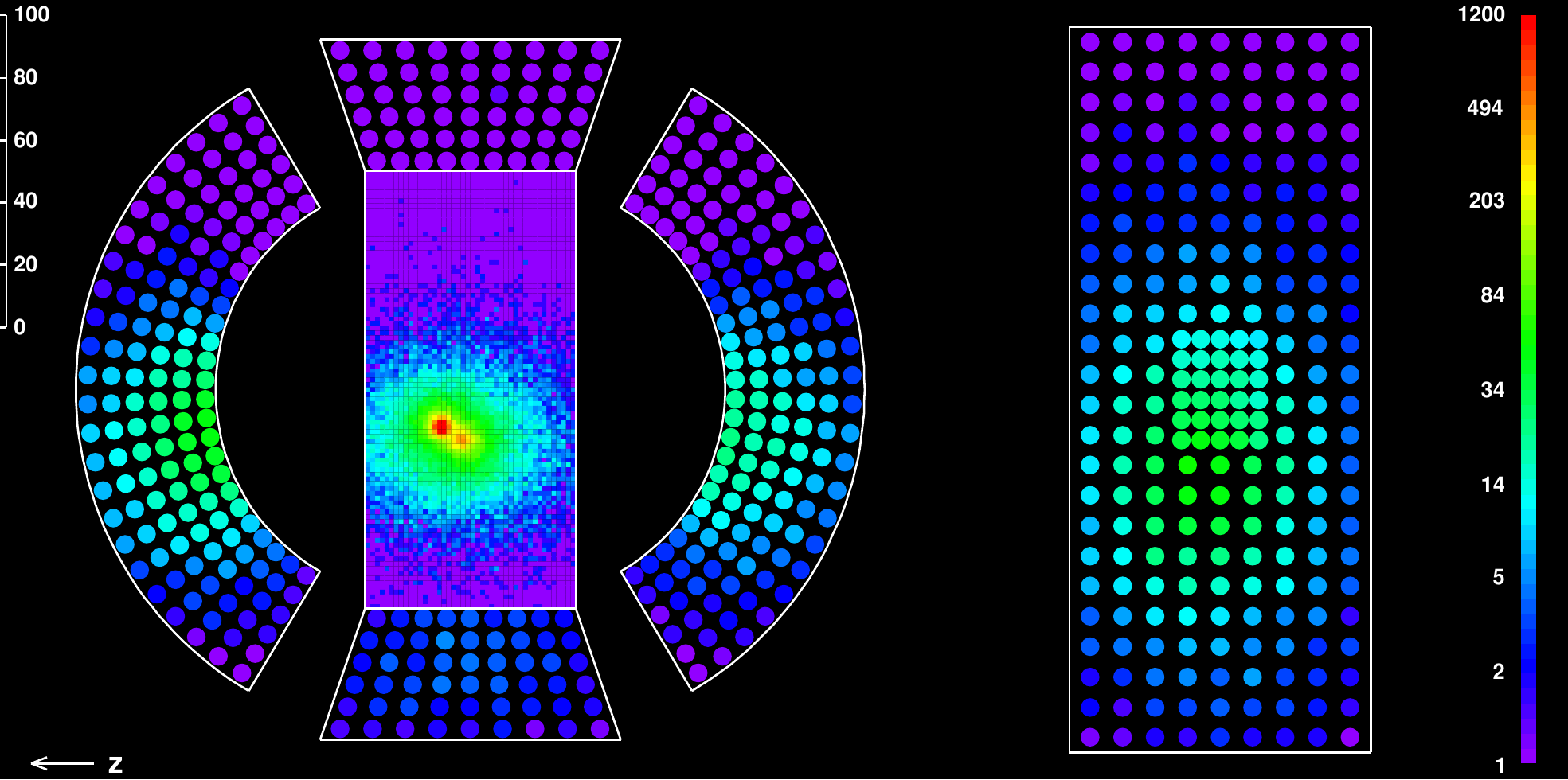}
\caption{\label{fig:comparison_of_imaging}%
   Example of scintillator light distributions seen by photo-sensors
      in case of (left) PMTs and (right) smaller photo-sensors ($12\times 12~\mathrm{mm}^2$)
      on the photon entrance face against the same MC event.
}
\end{figure*}

The LXe detector has been retained as a photon detector (see Fig~.\ref{fig:LXeDetector}). It has been upgraded substituting the PMTs reading the front face with SiPMs.
This brings important improvements in photon position and timing resolution and in resolving closely spaced photons as shown in Fig.~\ref{fig:comparison_of_imaging}.

%

\begin{figure}[tb]
  \centering
  \includegraphics[width=1.0\textwidth]{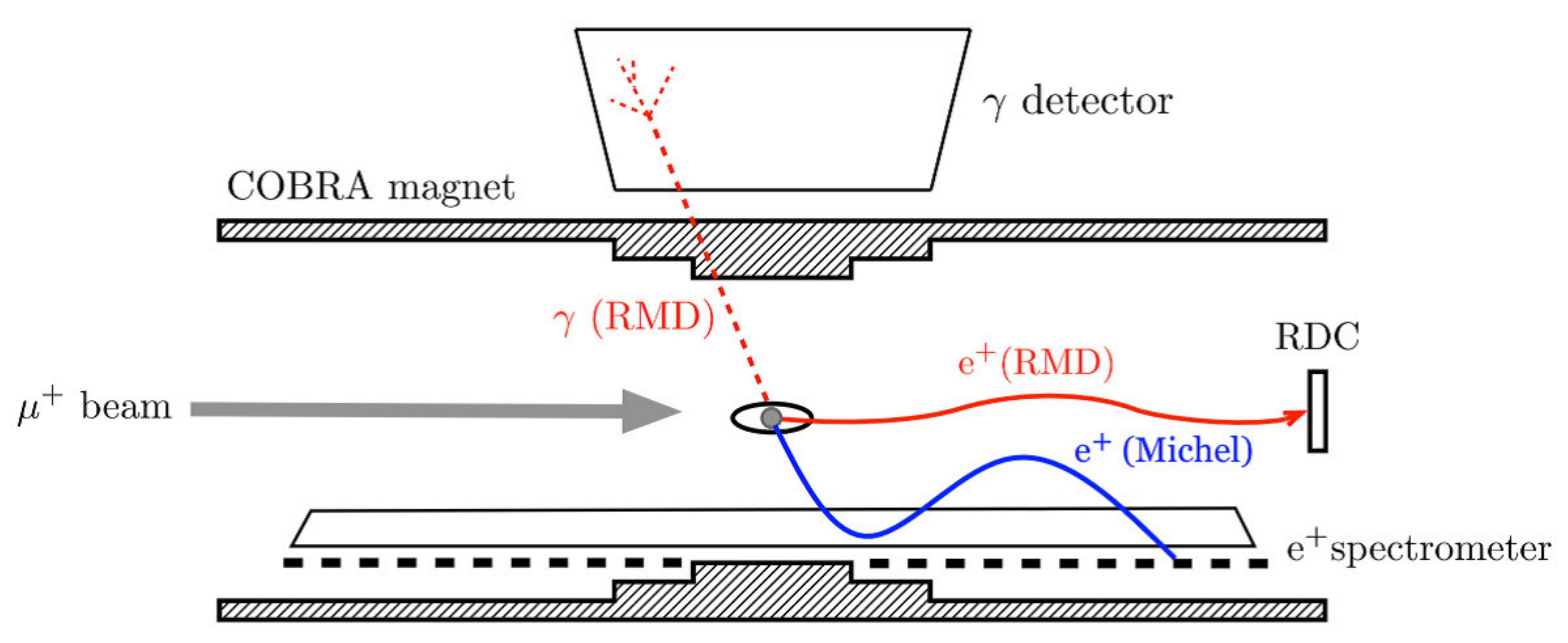}
  \caption{
  \label{fig:rdc-sketch}%
   Schematic view of the detection of RMD with the RDC detector.
  }
\end{figure}

A new subdetector added to MEG~II is the downstream RDC detector.
It is capable of identifying a fraction of the RMD decays with the photon energy close to
the kinematic limit, which are the dominant source of background photons for accidental coincidences. 
The basic concept is depicted in Fig.~\ref{fig:rdc-sketch}. The detector consists of a plane of plastic
scintillator plates for position measurement followed by a calorimeter based on LYSO crystals for energy 
measurements, both detectors are read out by SiPMs.

\begin{figure}
\centering
\includegraphics[width=1.0\linewidth, clip, trim=0 2pc 0 4pc]{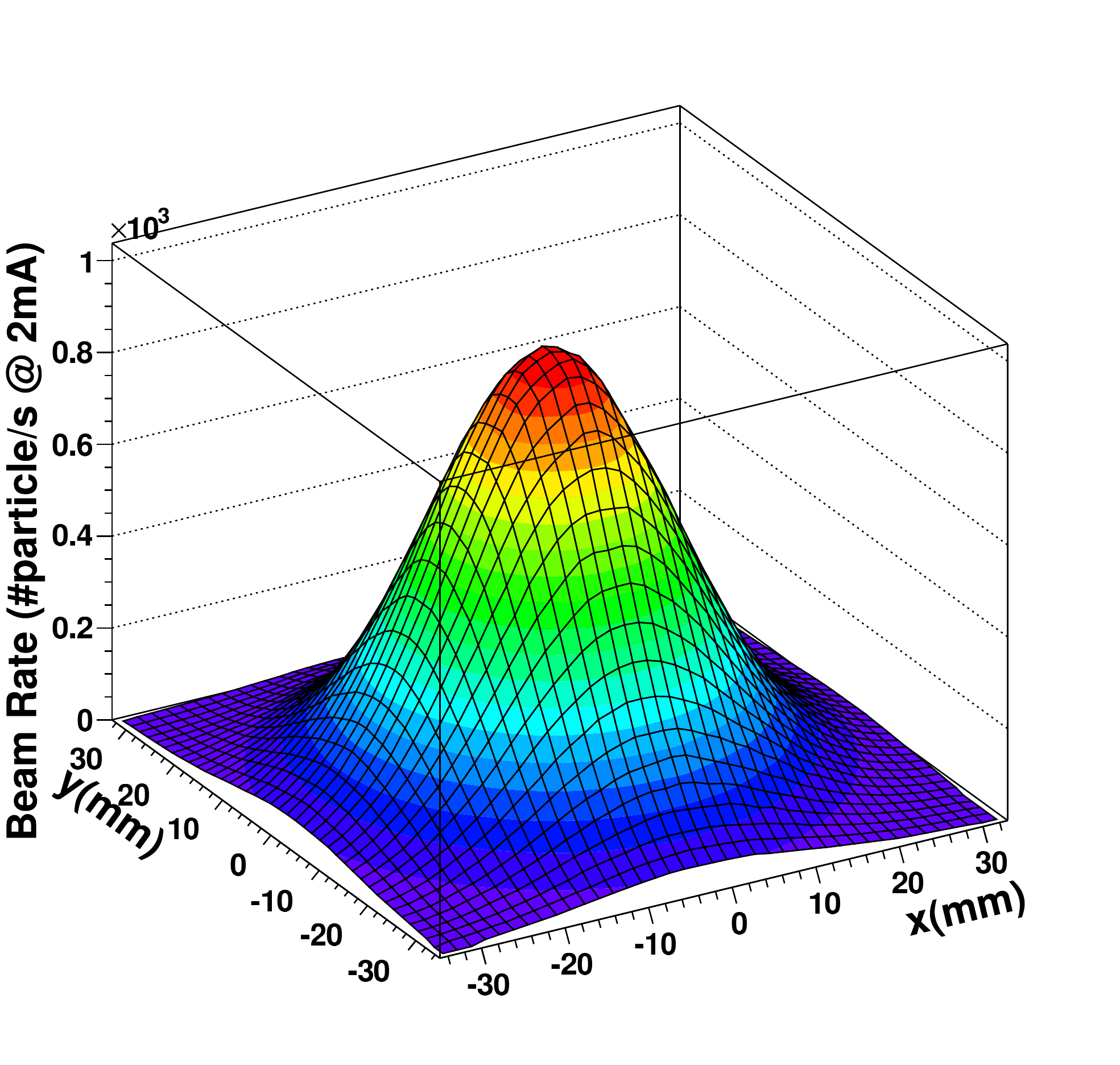}
\caption{Positive muon beam profile and rate as measured along the $\pi$E5 beam line.}
\label{fig:SciFiBeamProfiles}
\end{figure}

A crucial part of the experiment is the calibration and monitoring systems, some of them requiring dedicated hardware, 
to control over the lifetime of the experiment the responses of the subdetectors. In Table~\ref{tab:ListOfCalibrations} 
the list of calibration of tools to be used in MEG~II is presented. As an example, in Fig.\ref{fig:SciFiBeamProfiles} 
the result of a novel device for monitoring the beam spot based on scintillating fibers is presented.

\begin{table*}
\scriptsize 
\footnotesize 
\caption{\label{tab:ListOfCalibrations}%
The calibration tools of the MEG~II experiment.}
\centering
{\begin{tabular*}{\textwidth}{@{\extracolsep{\fill}}ccccc@{}}
\hline
\multicolumn{2}{c}{\bf Process}  & {\bf Energy} & {\bf Main Purpose} & {\bf Frequency} \\
\hline
Cosmic rays & atmospheric $\mu^{\pm}$ & Wide spectrum $\cal O$(GeV) & LXe--CDCH relative position & Annually \\
& & & CDCH alignment & \\
& & & LXe purity & On demand\\
Charge exchange & $\pi^- \mathrm{p}  \to \pi^0 \mathrm{n} $  & $55, 83, 129$~MeV $\gamma$ & LXe energy scale/resolution & Annually \\
& $ \pi^0 \to \gamma \gamma$ & & & \\
Radiative $\mu-$decay  & $\radiative$ & Photons $> 40$~MeV,   & LXe--pTC relative timing & Continuously \\
& & Positrons $> 45$~MeV & Normalisation & \\
Normal $\mu-$decay  & $\michel$ & 52.83 MeV & CDCH energy scale/resolution & Continuously \\
                    &           & end-point $\positron$s & CDCH energy scale/resolution & Continuously \\
& & & CDCH and target alignment & \\
& & & pTC time/energy calibration & \\
& & & Normalisation & \\
Mott positrons & $\positron$ target $\to \positron$ target & $\approx 50$~MeV $\positron$s & CDCH energy scale/resolution & Annually \\
& & & CDCH alignment & \\
Proton accelerator & $^7 {\rm Li} (\mathrm{p}, \gamma) ^8 {\rm Be}$ & 14.8, 17.6~MeV photons & LXe uniformity/purity& Weekly \\
& $^{11} {\rm B} (\mathrm{p}, \gamma) ^{12} {\rm C}$ & 4.4, 11.6, 16.1~MeV photons & LXe--pTC timing & Weekly \\
Neutron generator & $^{58} {\rm Ni}(\mathrm{n},\gamma) ^{59}{\rm Ni}$ & 9~MeV photons & LXe energy scale & Weekly \\
Radioactive source & $^{241}{\rm Am}(\alpha,\gamma)^{237}{\rm Np}$ & 5.5~MeV $\alpha$'s & LXe PMT/SiPM calibration & Weekly \\
& & & LXe purity& \\
Radioactive source & $^9\mathrm{Be}(\alpha_{^{241} {\rm Am}}, \mathrm{n})^{12}\mathrm{C}^{\star}$ & 4.4~MeV photons & LXe energy scale & On demand \\
                   & $^{12}\mathrm{C}^{\star}(\gamma)^{12}\mathrm{C}$ & & & \\
Radioactive source & $^{57}\mathrm{Co}(\rm{EC},\gamma)^{57}\mathrm{Fe}$ & 136 (11 $\%$), 122 keV (86 $\%$) & LXe--spectrometer alignment & Annually \\
                   &                                                    & X-rays & \\
LED & & UV region & LXe PMT/SiPM calibration & Continuously \\
Laser & &  401 nm  & pTC inter-counter timing & Continuously \\
\hline
\end{tabular*}}
\end{table*}

\begin{figure}
\centering
\includegraphics[width=1.0\textwidth, clip, trim=0 1pc 1pc 5pc]{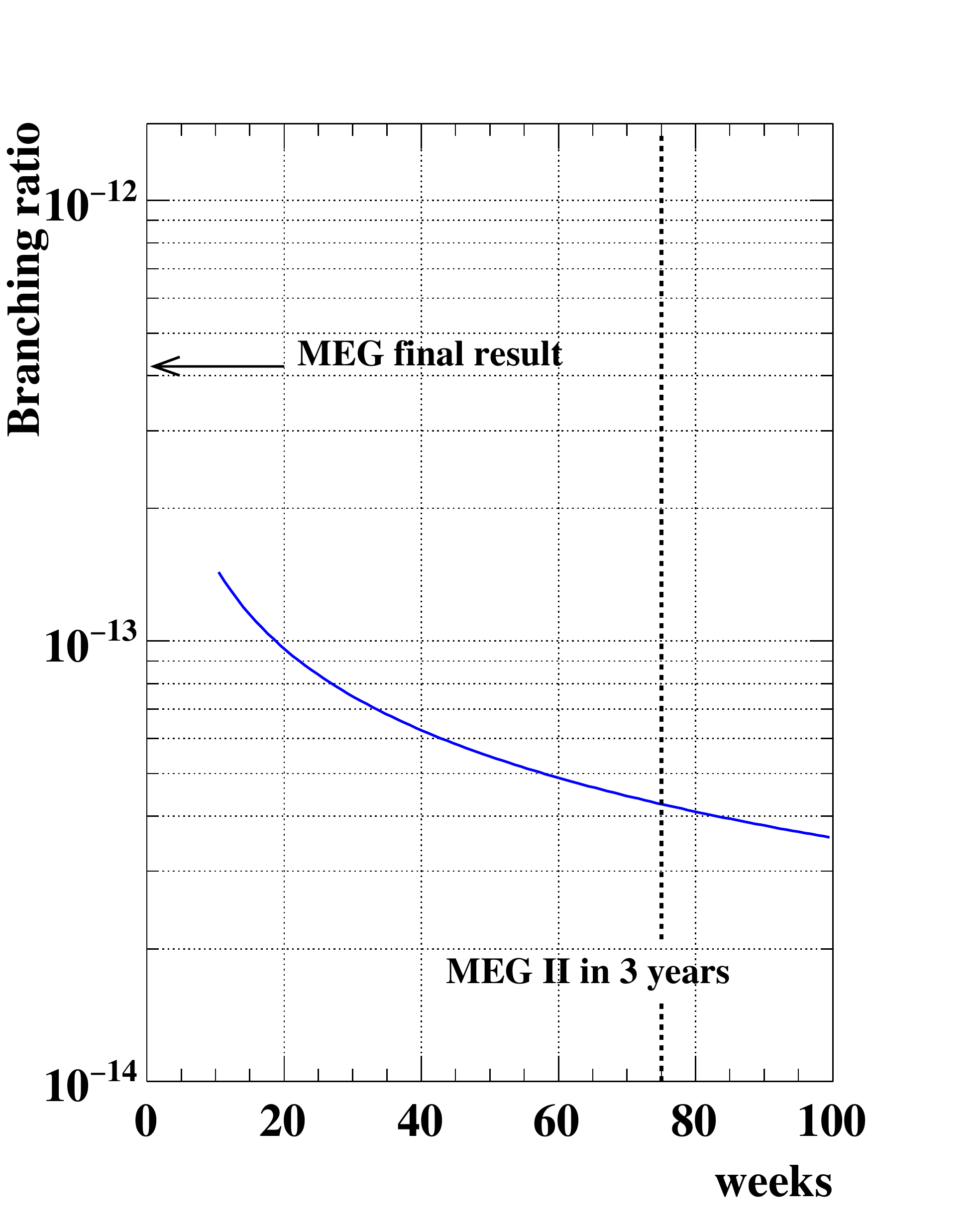}
\caption{\label{fig:sensEv}Expected sensitivity of MEG~II as a function of
the DAQ time compared with the bounds set by MEG \cite{meg2016}.
Assuming $25$ DAQ weeks per year, we expect a
90\%-CL UL  $\BR(\megc) < 4.3 \times 10^{-14}$
in three years.}
\end{figure}

\begin{table}
\caption{ \label{tab:scenario}Resolutions (Gaussian $\sigma$) and efficiencies of MEG~II compared
with those of MEG}
\centering
\newcommand{\m}{\hphantom{$-$}}
\newcommand{\cc}[1]{\multicolumn{1}{c}{#1}}
\begin{tabular}{@{}lll}
\hline
  {\bf PDF parameters }  & \m MEG & \m  MEG~II \\
\hline\noalign{\smallskip}
$\epositron$ (keV)                & \m  380    & \m 130   \\
$\thetae$ (mrad)                  & \m 9.4     & \m 5.3     \\
$\phie$ (mrad)                    & \m 8.7     & \m 3.7     \\
$\zpos/\ypos$ (mm) core            & \m 2.4/1.2 & \m 1.6/0.7  \\
$\egamma$(\%)  ($w>$2 cm)/($w<$2 cm))     & \m 2.4/1.7 & \m 1.1/1.0 \\
$\ugamma,\vgamma,\wgamma$ (mm)          & \m 5/5/6   & \m 2.6/2.2/5  \\
$\tegamma$ (ps)                   & \m 122     & \m 84 \\
\hline
{\bf  Efficiency (\%)} & &  \\
\hline
Trigger             & \m $\approx$ 99  & \m  $\approx$ 99 \\
Photon        & \m 63            & \m  69 \\
$\positron$               & \m 30            & \m  70  \\
\hline
\end{tabular}
\end{table}

In Table~\ref{tab:scenario} the resolutions and efficiencies required by the calculation of the sensitivity are 
presented for MEG (measured) and MEG`II (expected); the improvements are clear.\\
In Fig~\ref{fig:sensEv} the MEG~II sensitivity versus data taking time is shown. The MEG final limit ($4.3\,10^{-14}$) 
and the MEG~II expectation for three years of data taking are shown.

\acknowledgments

We are grateful for the support and co-operation provided
by PSI as the host laboratory and to the technical and
engineering staff of our institutes. 

\bibliographystyle{JHEP}
\bibliography{MEGIIINSTR17}
\end{document}